# A strain tensor that couples to the Madelung stress tensor

D. H. Delphenich
Kettering, OH 45440 USA

**Abstract.** Ordinarily, the stress tensor that one derives for a Madelung fluid is not regarded as being coupled to a strain tensor, which is consistent with the fluid hypothesis. However, based upon the author's earlier work regarding the geometric nature of the quantum potential, one can, in fact, define a strain tensor, which is not, however, due to a deformation of a spatial region, but to a deformation of a frame field on that region. When one expresses the Madelung stress tensor as a function of the strain tensor and its derivatives, one then defines a constitutive law for the Madelung medium that might lead to a more detailed picture of its elementary structure. It is pointed out that the resulting constitutive law is strongly analogous to laws that were presented by Kelvin and Tait for the bending and torsion of elastic wires and plates, as well as the Einstein equations for gravitation if one takes the viewpoint of "metric elasticity."

**1. Introduction.** Since the outset, the Schrödinger equation for the time evolution of matter waves was open to interpretation. Mostly, the issue was how to physically interpret the wave function $\Psi(t, x^i)$ that would constitute its solution for either an initial-value problem or a boundary-value problem relating to its stationary form. In particular, one needed to interpret the real scalar function $\rho = \|\Psi\|^2 = \Psi\Psi^*$.

Schrödinger himself originally regarded it as representing the electric charge density of the particle that was described by the wave function, but soon the Copenhagen School – principally, Bohr, Born, and Heisenberg – replaced that interpretation with the currently-accepted "statistical" interpretation. In that picture, $\rho$, when normalized to have total "mass" 1, would represent a probability density function that was associated with presence of a point-like particle within a given region of space and time by integration.

Not all of the theoretical physics community − including Schrödinger himself − were convinced that the statistical interpretation was the best way to resolve the issue. Einstein, who was actually quite distinguished in the statistical physics community, saw the introduction of stochastic contributions to a physical model as a fundamental incompleteness in the model. He famously said "God does not play dice!" in a letter to Max Born, to which Niels Bohr openly replied "Who are we to tell God what to do!"

One intriguing alternative to the statistical interpretation that emerged quite soon after Schrödinger introduced his equations was the so-called "hydrodynamical" interpretation, which basically originated with Ernst Madelung [**1**], and was expanded considerably by Takabayasi [**2**], who also discussed the relativistic form of the Madelung equations that came from the Klein-Gordon equation, as well as the hydrodynamical interpretation of the Dirac wave function. Actually, as we will see later, the use of the designation "hydrodynamical" is somewhat casual, since the stress tensor of a Madelung medium is not precisely consistent with the usual form of conventional fluid media. Since one is, after all, dealing with the structure of matter at the quantum (i.e., atomic to sub-atomic) level, it seems reasonable to be cautious in defining the nature of a quantum medium.

In the Madelung interpretation of the Schrödinger equation, the function $\rho$ is originally regarded as a number density for a particle distribution, which can be converted into a mass density by multiplying it by the mass $m$ of the particle, which then becomes the total mass associated with the density. Note that since $\rho$ takes on real values, and not just integer values, one must regard the non-integer values for its integral over a region of space and time as representing the *fraction* of the particle that exists within the region. It is then unnecessary to propose any "sub-quantum" level at which the particle resolves to more particles, as in the kinetic theory of gases.

By putting the wave function $\Psi$ into polar form $R\,e^{iS/\hbar}$, in which $R$ and $S$ are real functions of space and time, and substituting into the Schrödinger equation, upon separating out the resulting real and imaginary equations, one obtains a pair of partial differential equations that take the form of an equation of continuity for $\rho = R^2$ and a Hamilton-Jacobi-type equation for the energy of the wave. Since the only place in all of this where Planck's constant appears is in a potential energy term in the latter equation, one usually treats that potential energy as being of "quantum" origin.

Of course, that simply changes the nature of the interpretation problem from one of interpreting the wave function to one of interpreting the quantum potential. In an earlier paper by the present author [**3**], it was discovered that the Madelung potential could be given a geometric interpretation. Basically, the starting point was that if one used $\rho$ as a conformal factor to transform the Euclidian metric (in the non-relativistic case) or the Minkowski metric (in the relativistic case) then the Madelung potential was closely related to the scalar curvature of Levi-Civita connection that one obtains from the conformally-transformed metric. One notes that if $\rho$ is a mass density function then the conformally-transformed metric is what one uses to associate a momentum density 1-form with a velocity vector field.

The present effort succeeds in going a step beyond the previous attempt by showing that one can also define a measure of the strain that is associated with the deformation of the natural frame field on a region of space or spacetime by the dilatation that is defined by $R = \sqrt{\rho}$ such that the equation for the stress tensor that is associated with a Madelung medium (either relativistic or not) is essentially a first-order, nonlinear constitutive law that couples the stress associated with that deformation of the frames to the strain. What begins to emerge from this is the deeper idea that one must consider both the "kinematical" geometry of space, which is independent of the presence of matter, and the "dynamical" geometry, which is intimately related to the matter distribution. A crucial difference in methodology between the present work and the previous one is that, previously, the connection that was defined by the deformed frames was the Levi-Civita connection, whereas in the present investigation it was found that it is much more direct to use the "teleparallelism" connection; i.e., the connection that makes the deformed frame field parallel.

The general flow of ideas from here on is that in the next section, we shall summarize the basic definitions and results of the Madelung-Takabayasi program, while mostly focusing on the stress tensor that results from it. In section 3, we then propose a definition of frame strain that is closely-related to the Cosserat approach [**4**] to deformable media, which emphasizes the role of frame fields on the regions being deformed. The definition of infinitesimal frame strain that emerges from this is then shown, in section 4, to be the 1-form of the teleparallelism connection that is defined by the deformed frame field, and that approach to differential geometry is reviewed. In section 5, we then attempt to justify that mechanical constitutive laws for frame strain have been around for some time, and appeared as early as 1895 in Kelvin and Tait's treatise on natural philosophy [**5**] in their discussion of the bending and twisting of elastic wires. We also point out that one can regard Sakharov's notion of "metric elasticity" [**6**] as amounting to interpreting Einstein's equations of gravitation as a second-order constitutive law in the metric strain, or, if one represents the metric by "vierbeins," frame strain. We then conclude with a discussion of the results.

**2. The Madelung medium.** In order to discuss the physical properties of the Madelung medium, one must first address the Madelung form of the Schrödinger equation, and then introduce a Lagrangian density for those equations. From there, one can obtain the stress tensor for the medium. For the sake of completeness, we shall repeat the discussion in the relativistic case, where one applies the same transformation to the Klein-Gordon equation and obtains analogous results.

*a. The non-relativistic Madelung equations.* If one desires to obtain the Madelung equations for non-relativistic wave mechanics then one starts with the time-dependent Schrödinger equation:

$$\left[ \frac{\hbar}{i} \frac{\partial}{\partial t} - \frac{\hbar^2}{2m} \Delta + V \right] \Psi = 0, \tag{2.1}$$

and applies the Madelung transformation, which amounts to introducing polar coordinates into the field space $\mathbb{C}$:

$$\Psi = R e^{-iS/\hbar}. \tag{2.2}$$

in which:

$$\rho \equiv \| \Psi \|^2 = R^2. \tag{2.3}$$

The function $\rho(t, x)$ can be regarded as a number density for the matter distribution that is being described, while the function $S(t, x)$ can be regarded as the action function for the matter wave.

After separating the resulting complex equation into its real and imaginary parts, and defining the momentum density 1-form $p$, along with its vector field $\mathbf{p}$, one then gets the Madelung equations:

$$\left. \begin{aligned} 0 &= m\dot{\rho} + \operatorname{div}(\rho \mathbf{p}), \\ 0 &= \dot{S} + \frac{1}{2m} p^2 + V - \frac{\hbar^2}{2m} \frac{\Delta R}{R}, \\ p &\equiv dS = mv = i_{\mathbf{p}} \delta. \end{aligned} \right\} \tag{2.4}$$

The first equation in (2.4) takes the form of a continuity equation; thus, it describes the conservation of mass along the flow. From its form, one sees that the flow is generally compressible.

The second equation in (2.4) takes the form of a Hamilton-Jacobi equation, which is then associated with the balance of energy along the flow. In it, one sees that the only term in which the quantum constant $\hbar$ appears is a term that is sometimes called the *Madelung potential:*

$$V_q = -\frac{\hbar^2}{2m}\frac{\Delta R}{R}. \tag{2.5}$$

This then implies that one also can define a quantum force:

$$f_q = -dV_q = \frac{\hbar^2}{2m} d\left(\frac{\Delta R}{R}\right), \tag{2.6}$$

whose support is contained within the support of $\rho$ or $R$. Thus, it takes the form of an internal force that acts upon the mass distribution.

Note that a constant density would produce no quantum force, so the role of quantum force seems to be related to maintaining the inhomogeneity of the density. This suggests that one might be dealing with some process of diffusion whose equilibrium state is not a constant density, as is usually the case.

The last equation in (2.4) basically defines the momentum density 1-form $p$ of the object in motion in a manner that also suggests Hamilton-Jacobi notions. Because $p$ and the covelocity 1-form $v = p / m$ are both exact 1-forms they are both closed, as well. Hence, the kinematical and dynamical vorticities of the flow that is described by the Madelung equations both vanish:

$$0 = \Omega_k = dv, \qquad 0 = \Omega_d = dp. \tag{2.7}$$

Hence, the flow is irrotational.

*b. Non-relativistic Madelung stress tensor.* In order to derive a stress tensor for the Madelung medium, the simplest method is to start with a Lagrangian for the Schrödinger equation, convert it into a Lagrangian for the Madelung equations, and then derive the stress tensor from Noether's theorem.

A Lagrangian for the Schrödinger equation is:

$$\mathcal{L}(\Psi, \Psi^*, d\Psi, d\Psi^*) = \frac{i\hbar}{2}(\Psi^*\dot{\Psi} - \dot{\Psi}^*\Psi) - \frac{\hbar^2}{2m}<d\Psi^*, d\Psi> - V\Psi^*\Psi. \tag{2.8}$$

Applying the Madelung transformation (2.2) to this makes:

$$\mathcal{L}(\rho, S, d\rho, dS) = -\rho\left[\dot{S} + \frac{1}{2m}(dS)^2 + V + \frac{\hbar^2}{8m}\frac{(d\rho)^2}{\rho^2}\right]. \tag{2.9}$$

From this Lagrangian, one can obtain a stress tensor that takes the form (cf. [**2**, 1952], App. A):

$$\sigma_{ij} = \frac{\hbar^2}{4m}\rho\frac{\partial^2(\ln\rho)}{\partial x^i \partial x^j} = \frac{\hbar^2}{2m}\left[R\frac{\partial^2 R}{\partial x^i \partial x^j} - \frac{\partial R}{\partial x^i}\frac{\partial R}{\partial x^j}\right]. \tag{2.10}$$

This tensor is obtained from the canonical energy-moment-stress tensor that one gets from $\mathcal{L}$ by way of Noether's theorem when one subtracts the kinetic part that relates to $S$.

One notes that this tensor is symmetric, and can thus be diagonalized. The fact that there seem to be shearing stresses present suggest that it does not generally take the form of the stress tensor for a perfect fluid:

$$\sigma_{ij} = -\pi\, \delta_{ij}, \tag{2.11}$$

in which $\pi(t, x)$ is the pressure. In fact, this is true only when $\rho$ is a radially-symmetric, Gaussian density:

$$\rho(r) = \rho_0 e^{-ar^2}. \tag{2.12}$$

Thus, one sees that the use of the terms "fluid" or "hydrodynamics" in regard to the Madelung medium are somewhat premature. Since one is, after all, dealing with a state of matter that is found at the subatomic level, one should probably be more objective in identifying the precise nature of that state.

Since $\sigma_{ij}$ does not involve the velocity gradient, there is no viscosity present.

The equilibrium equation that one derives from (2.10) is:

$$\partial_j \sigma_i^j = \rho f_i, \tag{2.13}$$

in $f_i$ is the quantum force that is given by (2.6).

The stress tensor is associated with a mean pressure:

$$\bar{\pi} = -\tfrac{1}{3}\sigma_i^i = \frac{\hbar^2}{12m}\rho \Delta(\ln \rho) = \frac{\hbar^2}{6m}(\|\, dR\,\|^2 - R\, \Delta R). \tag{2.14}$$

This can be negative (i.e., a tension) or positive (i.e., the source of a dilatation) depending upon the nature of $R$. It vanishes iff $R\, \Delta R = \|\, dR\, \|^2$ or $\ln \rho$ is harmonic.

One can also express $\bar{\pi}$ in terms of the quantum potential (2.5):

$$\bar{\pi} = \tfrac{1}{3}\rho V_q + \frac{\hbar^2}{6m}\|\, dR\, \|^2. \tag{2.15}$$

If $\rho$ is a number density then the first term in the right-hand side of this expression represents an energy density. Since the second term is always positive, the only possible source of a negative mean pressure is the quantum potential.

   *c. Relativistic Madelung medium.* One can duplicate the process above by starting with the relativistic form of the Schrödinger equation that takes the form of the Klein-Gordon equation:

$$\Box\Psi + \left(\frac{m_0 c}{\hbar}\right)^2 \Psi = 0, \tag{2.16}$$

in which $\square = \eta^{\mu\nu} \partial_{\mu\nu}$ ($\eta^{\mu\nu} = \text{diag}[+1, -1, -1, -1]$) is the d'Alembertian operator.

Applying the same transformation as in (2.2) and separating out the real and imaginary equations gives the relativistic form of the Madelung equations:

$$\left. \begin{array}{l} 0 = \text{div}(\rho \mathbf{p}), \\ 0 = p^2 - m_0^2 c^2 + \hbar^2 \dfrac{\square R}{R}, \\ p \equiv dS = mv = i_{\mathbf{p}} \eta. \end{array} \right\} \tag{2.17}$$

This time, one can think of the second equation as saying that:

$$p^2 = m^2 c^2, \tag{2.18}$$

in which we have introduced:

$$m = m_0 \left( 1 - \frac{\hbar^2}{c^2} \frac{\square R}{R} \right)^{1/2}. \tag{2.19}$$

Thus, the point-like rest mass $m_0$ has been smeared out into a mass density. One can also say that the quantum potential energy deducts from the rest energy of the mass.

One can also put the second equation into the form of a modified dispersion law for the frequency-wave number 1-form $k = p/\hbar$:

$$k^2 = k_0^2 - \frac{\square R}{R}, \tag{2.20}$$

in which $k_0 = m_0 c / \hbar$ is the Compton wave number for the point-particle that is described by $\Psi$ when its rest mass is $m_0$.

Equations (2.17) can be derived from the Lagrangian density:

$$\mathcal{L} = -\frac{\hbar^2}{2m_0} \|dR\|^2 - \frac{1}{2} \rho \left( \frac{1}{m_0} \|dS\|^2 + m_0 c^2 \right). \tag{2.21}$$

The stress tensor that one derives from this is then (cf., [**4**, 1953], App. F):

$$\sigma_{\mu\nu} = \frac{\hbar^2}{4m_0} \rho \frac{\partial^2 \ln \rho}{\partial x^\mu \partial x^\nu} = \frac{\hbar^2}{2m_0} \left( \frac{\partial^2 R}{\partial x^\mu \partial x^\nu} - \frac{\partial R}{\partial x^\mu} \frac{\partial R}{\partial x^\nu} \right), \tag{2.22}$$

which compares quite closely with (2.10), since it differs by only the extra dimension of time. Consequently, the equilibrium equations that one must consider are really equations of motion:

$$\partial_\nu (m_0 \rho u^\mu u^\nu) = \partial_\nu \sigma^{\mu\nu}, \tag{2.23}$$

which we can put into the more traditional relativistic form:

$$\partial_\nu T^{\mu\nu} = 0, \tag{2.24}$$

if we define the energy-momentum-stress tensor in the obvious way:

$$T^{\mu\nu} = m_0\, \rho\, u^\mu u^\nu - \sigma^{\mu\nu}. \tag{2.25}$$

This time, when we compute the mean pressure, we must remember that there are now four diagonal components to $\sigma^{\mu\nu}$:

$$\bar\pi = -\tfrac{1}{4}\sigma^\mu_\mu = -\frac{\hbar^2}{16 m_0}\rho\,\Box(\ln\rho) = -\frac{\hbar^2}{8 m_0}(\|dR\|^2 - R\Box R). \tag{2.26}$$

**3. Frame strain.** In the author's previous study of the geometric origin of the quantum potential [**3**], it was discovered that everything seemed to originate from a very elementary transformation of the flat spatial metric, namely, the conformal transformation:

$$g_{ij} = \rho(t, x)\, \delta_{ij}, \tag{3.1}$$

in which we are describing the components with respect to a local frame field that is orthonormal for the Euclidian metric. If that frame field is the natural one $\partial_i = \partial/\partial x^i$ for some local coordinate chart $(U, x^i)$ then the Euclidian metric takes the form:

$$\delta = \delta_{ij}\, dx^i\, dx^j \tag{3.2}$$

for that frame field, in which $dx^i$ is the reciprocal coframe field to $\partial_i$, so $dx^i(\partial_j) = \delta^i_j$.

However, the approach taken at that time to describing the geometry of the deformed metric was the traditional Riemannian one of looking at the Riemann curvature of the Levi-Civita connection that was defined by the deformed metric. Since then, the author has found that there is a simpler approach to defining the strain that takes the metric $\delta$ to the metric $g = \rho\delta$ as it relates to the Madelung stress tensor in the form of the teleparallelism connection that is defined by the deformed frame field that is orthogonal for $g$. Thus, although we could continue in the Levi-Civita direction and obtain a coupling of strain to stress, we shall pursue the simpler path of teleparallelism.

*a. The deformation of a metric.* Given the conformal transformation (3.1), one can define a coframe field $\{\theta^i, i = 1, 2, 3\}$ that is orthonormal for the deformed metric, so:

$$g = \delta_{ij}\, \theta^i\, \theta^j, \tag{3.3}$$

by way of the coframe transformation:

$$\theta^i = y^i_j\, dx^j = R\, dx^i, \qquad y^i_j = \sqrt{\rho}\, \delta_{ij}, \tag{3.4}$$

in which the quantity $R = \sqrt{\rho}$ now plays a more fundamental role than $\rho$.

Now, in the Cauchy-Green conception of strain, the state of strain that is produced by a diffeomorphism of an extended material object is the difference between the deformed metric (pulled back to the initial object) and the initial metric. When the diffeomorphism is an isometry – i.e., a rigid motion – no strain will be produced. However, the transformation that we are dealing with at the moment is not generally due to a diffeomorphism of the object itself, but only a deformation of a frame field on it. In order for the transformation to be integrable into a diffeomorphism $y^i = y^i(x^j)$, which would make:

$$y^i_j = y^i_{,j} = \frac{\partial y^i}{\partial x^j}, \qquad (3.5)$$

the matrix $y^i_j$ would have to satisfy the necessary condition that:

$$y^i_{j,k} = y^i_{k,j}; \qquad (3.6)$$

i.e.:

$$d_\wedge \xi^i = 0, \qquad (3.7)$$

in which we have denoted the exterior derivative operator by $d_\wedge$ in order to not confuse it with the ordinary differential operator $d$.

However, in general, we will have:

$$d_\wedge \theta^i = dR \wedge dx^i = \frac{1}{R} dR \wedge \theta^i \equiv \omega \wedge \theta^i, \qquad (3.8)$$

in which we have defined the 1-form:

$$\omega = \frac{1}{R} dR = d(\ln R) = \frac{\partial (\ln R)}{\partial x^i} dx^i = \frac{1}{2} \frac{\partial (\ln \rho)}{\partial x^i} dx^i. \qquad (3.9)$$

Thus, the only way that this could vanish is if $\omega$ were collinear with all of the coframe members. However, since they are all linearly independent and span the cotangent spaces, moreover, this is only possible if $d\rho$ vanishes. That would make $\rho$ locally constant (i.e., constant on the connected components of the object). Hence, whenever $\rho$ is non-constant, one will be dealing with a frame transformation that does not integrate to a diffeomorphism of the object.

One notes that the differential of $\omega$ can be expressed in the local form:

$$d\omega = \frac{1}{2} \frac{\partial^2 (\ln \rho)}{\partial x^i \partial x^j} dx^i dx^j. \qquad (3.10)$$

One now sees that if we express our Madelung stress tensor in the form $\sigma = \sigma_{ij} dx^i dx^j$ then the first expression in (2.10) takes the elementary form:

$$\sigma = \frac{\hbar}{2m} \rho \, d\omega. \tag{3.11}$$

If we can physically justify regarding $\omega$ as a type of strain then we can regard this last equation as a first-order constitutive law that couples the strain on the metric that is induced by the local homothety to a corresponding stress (by way of its differential). One notes that no stress is produced when $\rho$ – or even its differential – are constant.

Another aspect of equation (3.11) that suggests that it represents a constitutive law is the appearance of $\hbar$ in it. At a more elementary level, one can regard the de Broglie relation:

$$p = \hbar k, \qquad p \equiv E\, dt + p_i\, dx^i, \qquad k \equiv \omega\, dt + k^i\, dx^i \tag{3.12}$$

as a mechanical constitutive law that couples the kinematical state of a wave, as defined by the frequency-wave number 1-form $k$, to the dynamical state, as defined by the energy-momentum 1-form $p$. Interestingly, if one considers a spatially-extended particle, so $p$ represents an energy-momentum density, and perhaps takes the form $\rho v$, where $p$ is the mass density and $v$ is the covelocity 1-form, then there is nothing to suggest that $\hbar$ is truly a constant. The fact that it is treated as such is due to the fact that quantum mechanics associates waves with *point-like* particles, so one must integrate the energy-momentum density over space in order to get the total energy-momentum 1-form along a curve.

At a more elementary level, one sees that we are associating the action function $S$ for the wave with its phase function $\theta$ by way of:

$$S = \hbar\, \theta, \tag{3.13}$$

which then gives (3.12) upon differentiation, but only as long as $\hbar$ is a constant.

*b. Frame strain.* In order to justify that (3.11) represents a mechanical constitutive law in which $\omega$ represents an infinitesimal strain, we first note that we are not deforming a region of space, but a frame field on a region, and that such a frame field deformation can still bring about the deformation of a metric. More generally, such a deformation can be produced by a set of equations of the form:

$$\theta^i = h^i_j(x)\, dx^j, \tag{3.14}$$

in which the matrix $h^i_j(x)$ is invertible for every $x$.

If one factors the matrix $h(x)$ into the product $RE$ of a rotation $R$ with a finite strain $E$ at each $x$ then the deformed metric will have a component matrix:

$$[g] = E^\mathrm{T} R^\mathrm{T} R E = E^\mathrm{T} E. \tag{3.15}$$

Thus, the only part of the matrix $h$ that affects the metric is the matrix $E$, which can then be used as a measure of finite strain. In the case at hand of a Madelung medium, we have:

$$E^i_j = R\,\delta^i_j. \tag{3.16}$$

From the fact that:

$$d\theta^i = -\,\omega^i_j \otimes \theta^j, \tag{3.17}$$

in which ([1]):

$$\omega^i_j = dh^i_k\,\tilde{h}^k_j, \tag{3.18}$$

one sees that when one introduces the polar decomposition of $h$ as $RE$ this makes:

$$\omega^i_j = dR^i_k\,\tilde{R}^k_j + R^i_k\,dE^k_l\,\tilde{E}^l_m\,\tilde{R}^m_j. \tag{3.19}$$

One sees that the first term in the right-hand side of this expression represents a 1-form with values in the Lie algebra $\mathfrak{so}(3)$, in the Euclidian case, and thus represents infinitesimal rotations, while the second term takes its values in a vector space that is complementary to $\mathfrak{so}(3)$ in $\mathfrak{gl}(3)$ and represents infinitesimal strains.

In the present case, there is no rotation of the natural frame field, so $R = I$ and with $E$ as in (3.16), we get:

$$\omega^i_j = dE^i_k\,\tilde{E}^k_j = d(\ln R)\,\delta^i_j. \tag{3.20}$$

Hence, we are indeed justified in regarding the 1-form $\omega^i_j$ as a way of measuring the infinitesimal strain in the frame $dx^i$ that is created by the finite strain $E^i_j = h^i_j = R\,\delta^i_j$.

Having resolved that issue, we next shall discuss the geometrical nature of the 1-form $\omega$, and then the work that has been done before in theoretical mechanics that is closely related to constitutive laws of the form that we have defined above.

**4. Teleparallelism connection.** If the conformal transformation (3.1) of $\delta_{ij}$ to $g_{ij}$ takes the form of a local homothety of the tangent spaces, such that we can define a new coframe $\theta^i$ by way of (3.4), then the deformed metric can then be expressed in two ways:

$$g = g_{ij}\,dx^i\,dx^j = \delta_{ij}\,\theta^i\,\theta^j. \tag{4.1}$$

Thus, the deformed coframe is orthonormal for the deformed metric.

Now, if one considers the differential of the coframe field $\theta^i$ then one can put it into the forms:

$$d\theta^i = dR \otimes dx^i = -\,\omega \otimes \theta^i, \tag{4.2}$$

in which the 1-form $\omega$ is again the one defined in (3.9).

This last equation suggests that we can regard $\omega$ as a connection 1-form that takes its values in the Lie algebra $\mathbb{R}$, which contains the infinitesimal generators of (one-parameter families of) homotheties. We express its components with respect to the natural coframe field in either of two forms:

$$\omega_i = -\partial_i R, \qquad \omega_{ki}^j = -\partial_i R \; \delta_k^j . \qquad (4.3)$$

---

([1]) Here, and in the sequel, a tilde over a matrix will denote its inverse.

When (4.2) is rewritten in the form:

$$0 = \nabla \theta^i \equiv d\theta^i + \omega \otimes \theta^i, \qquad (4.4)$$

one sees that if $\omega$ is indeed a connection then it apparently makes the coframe field $\theta^i$ parallel. In fact, this is how one defines the *teleparallelism connection* ([1]) that is associated with a given local frame field: It is the connection on the (local) bundle of linear frames that makes the chosen frame field parallel (and therefore its reciprocal coframe field, as well).

Since a connection form is not tensorial, but obeys an inhomogeneous transformation law, one finds that the 1-form that corresponds to $\omega$ relative to the deformed coframe is zero. That is, if $\omega = \omega_i' \theta^i$ then:

$$\omega_i' = R \, \omega_i R^{-1} + dR \, R^{-1} = 0. \qquad (4.5)$$

Clearly, this can be the source of much confusion, since one must often establish which form of the connection 1-form $\omega$ one needs to use, namely, $\omega = d(\ln R)$, which relates to its components with respect to $dx^i$, or $\omega = 0$, which relates to its components with respect to $\theta^i$.

Actually, there is nothing special about the coframe field $\theta^i$ as far as $\omega$ is concerned, since one sees immediately that any other coframe field of the form $A_j^i \theta^j$ will give the same $\omega$, as long as the invertible matrix $A_j^i$ is constant as a function of spatial position. This amounts to the statement that a covector field is *parallel* with respect to $\theta^i$ iff its components with respect to that coframe field are constants.

One can show this by calculation if one starts with the fact that if $\alpha = \alpha_i \theta^i$ is a covector field then:

$$d\alpha = d\alpha_i \otimes \theta^i + \alpha_i \, d\theta^i = d\alpha_i \otimes \theta^i, \qquad (4.6)$$

since $\omega = 0$ when one is looking at components with respect to $\theta^i$. Thus, $d\alpha$ vanishes iff the components $\alpha_i$ are constant.

If one uses components with respect to $dx^i$, so $\alpha = \alpha_i' dx^i$, with $\alpha_i' = R\alpha_i$, this time, then:

$$d\alpha_i' = dR \, \alpha_i + R \, d\alpha_i = (-\omega\alpha_i + d\alpha_i) \, R, \qquad (4.7)$$

and $\alpha$ is parallel with respect to the natural frame field iff:

---

([1]) The author has compiled an anthology [**7**] of English translations of many of the early articles on Einstein's attempt to use the geometry of parallelizable manifolds – i.e., teleparallelism – to unify the theories of gravitation and electromagnetism.

$$0 = \nabla \alpha_i = d\alpha_i - \omega \alpha_i. \tag{4.8}$$

The deformed coframe does not have to be holonomic, since one has:

$$d_\wedge \theta^i = \frac{1}{2}\left(\frac{\partial R}{\partial x^j}\delta^i_k - \frac{\partial R}{\partial x^k}\delta^i_j\right) dx^j \wedge dx^k = \frac{1}{2\rho}\left(\frac{\partial R}{\partial x^j}\delta^i_k - \frac{\partial R}{\partial x^k}\delta^i_j\right)\theta^j \wedge \theta^k. \tag{4.9}$$

In the first form, one sees that the components of the "anholonomity" 2-form $d_\wedge \theta^i$ take the form of the negative of (twice) the anti-symmetric part of the connection 1-form $\omega$, when one uses the second form in (4.3); i.e.:

$$\omega^i_{jk} - \omega^i_{kj} = -\left(\frac{\partial R}{\partial x^j}\delta^i_k - \frac{\partial R}{\partial x^k}\delta^i_j\right). \tag{4.10}$$

In fact, since $\omega = 0$ when one is looking at things in the deformed coframe, the pull-down of the Cartan structure equation for the torsion 2-form of the linear connection $\omega$ on the bundle of linear frames by the deformed coframe is simply:

$$\Theta = d_\wedge \theta^i = \tfrac{1}{2}\Theta^i_{jk}\theta^j \wedge \theta^k. \tag{4.11}$$

That is:

$$\Theta^i_{jk} = -(\omega^i_{jk} - \omega^i_{kj}). \tag{4.12}$$

One can also characterize the $d_\wedge \theta^i$ independently of $\omega$ by means of the structure functions $c^i_{jk}(x)$ of $\theta^i$, which are defined by:

$$d_\wedge \theta^i = -\tfrac{1}{2}c^i_{jk}\theta^j \wedge \theta^k. \tag{4.13}$$

Thus:

$$\Theta^i_{jk} = -c^i_{jk}. \tag{4.14}$$

The curvature 2-form that one gets for $\omega$ from the Cartan structure equations clearly vanishes:

$$\Omega = d\omega + \omega \wedge \omega = 0. \tag{4.15}$$

Indeed, this is typical of the teleparallelism connection, more generally. That is, the vanishing of curvature for a connection on a region of space is a necessary condition for the existence of a local parallel frame field on that same region.

If one looks at the covariant derivative of the deformed metric with respect to the deformed connection then one finds that it also vanishes:

$$\nabla g = d\delta_{ij} \otimes \theta^j \theta^k = 0. \tag{4.16}$$

Thus, the deformed connection is a metric connection for the deformed metric.

It also preserves the deformed volume element that is defined by $\theta^i$:

$$V = \theta^1 \wedge \theta^2 \wedge \theta^3 = \frac{1}{3!} \varepsilon_{ijk}\, \theta^i \wedge \theta^j \wedge \theta^k, \tag{4.17}$$

since, once again, its components with respect to $\theta^i$ are constants.

Putting all of the pieces together, one see that what the teleparallelism connection allows one to do is "anholonomic Euclidian geometry," in the sense that relative to the anholonomic frame field that defines parallelism, geometry looks Euclidian, except for the torsion, which implies a translation of a vector when it is parallel-translated around a loop. Thus, one has "parallelism without parallelograms," as some characterized the situation in the early history of that kind of geometry.

Although we have concentrated on the Euclidian case, there is no fundamental obstacle to doing the same thing in four-dimensional Minkowski space $\mathfrak{M}^4 = (\mathbb{R}^4,\, \eta_{\mu\nu})$ when one starts with the conformal transformation:

$$g_{\mu\nu} = \rho\, \eta_{\mu\nu}. \tag{4.18}$$

One generally obtains expressions that are analogous to the ones above by a change of notation for the indices and the replacement of $\delta_{ij}$ with $\eta_{\mu\nu}$.

**5. Constitutive laws for frame strain.** As mentioned above, a metric $g$ that has the same form as (3.1) plays a fundamental role in mechanics, since it allows one to associate a momentum density covector (i.e., 1-form) $p$ with the velocity vector field **v** by way of:

$$p = i_\mathbf{v} g, \qquad (\text{i.e., } p(\mathbf{w}) = g(\mathbf{v}, \mathbf{w})) \tag{5.1}$$

If the components are described with respect to the natural frame then they are related by:

$$p_i = g_{ij}\, v^j = \rho\, v_i, \tag{5.2}$$

in which $v_i = \delta_{ij}\, v^j$ are the components of the covelocity 1-form $v = v_i\, dx^i$. The main difference is that $\rho$ represents the mass density, this time. Of course, if $m$ represents the total mass of the Madelung object and $\rho$ is the number density then $m\rho$ will represent a mass density. One would have to make it a specific postulate that $m\rho$ did, in fact, represent the true mass density, rather than some other function that had the same integral $m$ over all of space.

This relationship then represents a type of mechanical constitutive law, since it associates a dynamical state – viz., $p$ – with a kinematical one – viz., **v**. It also involves the spatial metric in a fundamental way. Hence, we shall refer to the metric $g$ as a *dynamical metric* on space, while the original metric (whether Euclidian or Lorentzian) will be thought of as a *kinematical metric*. The essential difference is then the fact that a kinematical metric is purely geometrical and independent of the physical nature of any material object that occupies a region of space, while the dynamical metric includes empirical information, such as a mass density, that must be specified.

In continuum mechanics, the kinematical state of the deformation of a region of space is defined by the strain that the deformation produces, while the dynamical state is defined by the stress distribution that results. Thus, we can say that in (3.11) we indeed have a mechanical constitutive law for the Madelung medium that couples the stress to the infinitesimal strain, as measured by $\omega$, we also see that it is of first order in $\omega$. Furthermore, it has the property that a constant dilatation would produce no stress, so the origin of the stress is entirely contained within the inhomogeneity of the density function $\rho$.

This latter fact has a certain physical reasonableness to it, if one considers the way that things work in diffusion processes. Generally, an inhomogeneous density of a gas or concentration of a solute tends to bring about diffusion (e.g., as a result of a chemical potential) in such a way that the equilibrium state becomes one of homogeneity. Thus, the stress that is produced by our metric deformation is somewhat analogous to the chemical potential that makes a homogeneous density represent an equilibrium state.

Of course, the immediate difference here is that $\sigma_{ij}$ can still vanish for an inhomogenous $\rho$, since it is of second order in that function, not first order. Thus, a constant gradient will also produce no stress.

Constitutive laws of the form in question have been around for some time already. In the monumental treatise of Kelvin and Tait [**5**] on natural philosophy, one finds that they address the bending and torsion of an elastic wire, which is described by a differentiable curve $x(s)$ in space, by coupling the couple-stresses that are produced to the matrix of functions $\omega_i^j$ that belongs to $\mathfrak{so}(3)$ and comes from the equation of a deformed orthonormal frame field $\mathbf{e}_i(s)$ along the curve:

$$\frac{d\mathbf{e}_i}{ds} = \omega_i^j \mathbf{e}_j. \tag{5.3}$$

If the frame field $\mathbf{e}_i(s)$ is adapted to the curve – so for instance, $\mathbf{e}_1(s)$ is the (unit) velocity vector field on the curve – then the anti-symmetric matrix $\omega_i^j$ takes the form:

$$\omega_i^j = \begin{bmatrix} 0 & -\lambda & \kappa \\ \lambda & 0 & -\tau \\ -\kappa & \tau & 0 \end{bmatrix}, \tag{5.4}$$

in which $\kappa$ and $\lambda$ are the *curvatures* of the curve in the $\mathbf{e}_1$-$\mathbf{e}_3$ and $\mathbf{e}_1$-$\mathbf{e}_2$ planes, while $\tau$ is the *torsion* of the curve in the $\mathbf{e}_2$-$\mathbf{e}_3$ plane. They then represent infinitesimal rotations about the *x*, *y*, and *y* axes, respectively. Indeed, one can express the components of $\omega_i^j$ by using the fact that since $\mathbf{e}_i(s)$ is an orthonormal frame field, one has:

$$\frac{d\mathbf{e}_i}{ds} = \sum_j <\dot{\mathbf{e}}_i, \mathbf{e}_j> \mathbf{e}_j, \tag{5.5}$$

which makes:

$$\kappa = - <\dot{\mathbf{e}}_1, \mathbf{e}_2> = + <\mathbf{e}_1, \dot{\mathbf{e}}_2>, \tag{5.6}$$

$$\lambda = + <\dot{\mathbf{e}}_1, \mathbf{e}_3> = - <\mathbf{e}_1, \dot{\mathbf{e}}_3>, \tag{5.7}$$

$$\tau = - <\dot{\mathbf{e}}_3, \mathbf{e}_2> = - <\mathbf{e}_3, \dot{\mathbf{e}}_2>. \tag{5.8}$$

$$0 = <\dot{\mathbf{e}}_i, \mathbf{e}_i> \quad \text{for all } i. \tag{5.9}$$

These are essentially equations (8) in v. II, § 614 of Kelvin and Tait. One notes that the first two are of second-order in $x^i(s)$, since $\dot{\mathbf{e}}_1 = \dot{\mathbf{v}}$ is the acceleration vector when one gives the curve arc-length (i.e., unit-speed) parameterization.

In § 595 of the same volume, they essentially define the vector $\boldsymbol{\kappa} = (\kappa, \lambda, \tau)$ to represent the state of strain for the deformed wire, which is equivalent to using $\omega_i^j$, which is the adjoint matrix to the vector $\boldsymbol{\kappa}$ relative to the Lie algebra on $\mathbb{R}^3$ that is defined by the cross product. They then propose a linear constitutive law in order to couple the strain to the couple-stresses (i.e., bending and twisting moments), which can be assembled into the covector $M = (K, L, T)$, namely:

$$M_i = A_{ij}\, \kappa^j. \tag{5.10}$$

This is the essence of equation (3) in that section.

Note that this older usage of the terms "curvature" and "torsion," which goes back to Frenet and Serret, is actually more precise in the context of the bending and twisting of wires than the newer Riemannian usage of those terms, which has more to do with the obstructions to integrability of parallel translation.

If one further confers equation (4) in § 644 of Kelvin and Tait (*loc. cit.*) then one will find that when one deforms a planar plate into a curved surface that can be described by a function $z(x, y)$, one physically reasonable way of coupling the couple-stress $(K, \Lambda, \Pi)$ to the deformation is:

$$\left. \begin{array}{l} K = A \dfrac{\partial^2 z}{\partial x^2} + c \dfrac{\partial^2 z}{\partial y^2} + b \dfrac{\partial^2 z}{\partial x \partial y}, \\[6pt] \Lambda = c \dfrac{\partial^2 z}{\partial x^2} + B \dfrac{\partial^2 z}{\partial y^2} + a \dfrac{\partial^2 z}{\partial x \partial y}, \\[6pt] 2\Pi = b \dfrac{\partial^2 z}{\partial x^2} + a \dfrac{\partial^2 z}{\partial y^2} + C \dfrac{\partial^2 z}{\partial x \partial y}. \end{array} \right\} \tag{5.11}$$

However, this set of equations couples the couple-stress to the partial derivatives of the function that defines the surface, while equation (5.10) is basically a constitutive law that couples the couple-stress matrix $M_j^i$ ( = the adjoint of the covector $M$) to the matrix $\omega_i^j$ in a linear and algebraic way. Moreover, one can see that the main difference between this infinitesimal rotation matrix $\omega_i^j$ and the teleparallelism connection matrix $\omega_{jk}^i$ is simply a matter of differentiating with respect to more than one parameter. Thus, one suspects that there is no reason to not also regard $\omega_{jk}^i$ as a measure of the infinitesimal strain in a frame field that has been deformed by $h^i_j$ and define the couple-stress matrix $M^{ik}_j$ accordingly. Basically, it takes its values in the dual vector space $\mathfrak{so}(3)^*$ of the Lie algebra $\mathfrak{so}(3)$, and when one evaluates the linear functional that is defined by the values of the vector field $M^i_j = M^{ik}_j \mathbf{e}_k$ on the 1-form $\delta \omega^j_i$, which represents a virtual displacement (i.e., variation) of the strain state, the result is the virtual work $\delta W$ that is required in order to accomplish that.

The methods of Kelvin and Tait, which are based in the deformation of frames, along with regions of space, were expanded considerably by the Cosserat brothers in a landmark treatise [**4**], although we shall not go further in that direction at the moment.

If one takes the position of Sakharov [**6**] that general relativity represents a type of "metric elasticity" then one can think of the Einstein equations, which couple the Lorentzian metric tensor field to the energy-momentum-stress tensor $T_{\mu\nu}$, namely:

$$T_{\mu\nu} = R_{\mu\nu} - \tfrac{1}{2} R\, g_{\mu\nu}, \qquad (5.12)$$

as representing a second-order constitutive law in the "metric strain," which we now briefly discuss.

One thinks of $g_{\mu\nu}$ as being the deformation of the flat Minkowski space metric $\eta_{\mu\nu} =$ diag$[+1, -1, -1, -1]$ by a finite strain tensor $E_{\mu\nu}$ (in the Cauchy-Green sense):

$$g_{\mu\nu} = \eta_{\mu\nu} + E_{\mu\nu}, \qquad (5.13)$$

so

$$\partial_\lambda g_{\mu\nu} = \partial_\lambda E_{\mu\nu}, \qquad (5.14)$$

which allows one to express the components of the Levi-Civita connection in terms of the metric strain:

$$\Gamma^\lambda_{\mu\nu} = \tfrac{1}{2} g^{\lambda\alpha}(\partial_\mu E_{\alpha\nu} + \partial_\nu E_{\mu\alpha} - \partial_\alpha E_{\mu\nu}). \qquad (5.15)$$

One can also express $g^{\mu\nu}$ in the form $\eta^{\mu\nu} + E^{\mu\nu}$, but the matrix $E^{\mu\nu}$ does not have to represent the inverse to the matrix $E_{\mu\nu}$, which might very well be non-invertible, to begin with. However, from the fact that $g^{\mu\nu}$ is the inverse of $g_{\mu\nu}$, the two matrices $E_{\mu\nu}$ and $E^{\mu\nu}$ must be related by:

$$0 = \eta_{\nu\kappa} E^{\kappa\mu} + E_{\nu\kappa} \eta^{\kappa\mu} + E_{\nu\kappa} E^{\kappa\mu}, \qquad (5.16)$$

so, up to first order, one has:

$$E^{\mu\nu} = -\eta^{\mu\kappa} \eta^{\nu\lambda} E_{\kappa\lambda}. \qquad (5.17)$$

Since the Riemann curvature tensor, as well as its contractions to the Ricci curvature tensor $R_{\mu\nu}$ and the scalar curvature $R$, are all obtained from the connection 1-form $\Gamma^\mu_\nu = \Gamma^\mu_{\nu\lambda} dx^\lambda$ by differentiation, one sees that equation (5.12) can be regarded as a constitutive law that is either of first order in $\Gamma^\mu_\nu$ or of second order in $E_{\mu\nu}$.

A significant difference between the Einstein tensor:

$$G_{\mu\nu} = R_{\mu\nu} - \tfrac{1}{2} R\, g_{\mu\nu} \tag{5.18}$$

on the right-hand side of (5.12) and the right-hand side of the Madelung stress expression (2.10) is that Einstein chose the form of the expression $G_{\mu\nu}$ specifically for the purpose of making $\partial_\nu T^\nu_\mu$ vanish, while, as we have seen, the divergence of the Madelung stress tensor does not vanish. However, when one includes the kinetic contribution, as we did above in (2.25), one does obtain a total energy-momentum-stress tensor with vanishing divergence.

One can also regard Einstein's equations as something that relates to frame strain, rather than metric strain, by introducing vierbeins. They basically define a Lorentzian coframe field $\theta^\mu = h^\mu_\nu dx^\nu$ on a region of spacetime, so the metric can be represent in the form:

$$g = \eta_{\mu\nu}\, \theta^\mu\, \theta^\nu. \tag{5.19}$$

Since this means that the components of $g$ can be expressed in terms of the components of the vierbein field by way of:

$$g_{\mu\nu} = \eta_{\kappa\lambda}\, h^\kappa_\mu\, h^\lambda_\nu, \tag{5.20}$$

this means that one can then regard $\theta^\mu$ as as deformation of the natural coframe field $dx^\mu$, in such a way that the Levi-Civita connection and Riemann curvature tensors can be expressed in terms of the vierbein field, rather than the metric. Hence, Einstein's equations become a constitutive law for the frame strain thus defined.

It should be pointed out that Einstein's equations can also be formulated in terms of the teleparallelism connection defined by the vierbein field, rather than the Levi-Civita connection. In that formulation, the torsion plays the role that curvature did in the Riemannian picture.

**5. Discussion.** Although we have made some progress in the foregoing towards discovering the internal structure of quantum waves, nonetheless, we still have an incomplete model in that we still need to account for the constitutive law for quantum stress and strain in terms of first principles and a fundamental model for the wave. Since the picture that one considers in the hydrodynamical representation does not seem to include a fundamental interaction that would be responsible for the inhomogeneity of the mass distribution, one might reconsider the electric interpretation of that density, which was also successfully reprised by Pauli and Weisskopf in their theory of mesons.

Here, one must recall that since quantum physics originated in the behavior of sub-atomic charges and photons, the Schrödinger – or even Klein-Gordon – equation is itself an incomplete description of electrons or photons. For one thing, it does not account for the spin (or really, magnetic moment) of the electron, which actually generates a powerful magnetic field near the electron, and which might contribute to the equilibrium configuration for the density function.

Furthermore, one must consider that the Schrödinger equation is a linear equation that does not account for the spatial localization of the support of the quantum wave function. Perhaps, one might consider nonlinear extensions of that equation or the Klein-Gordon equation as a closer approximation to the true constitutive law.

## References (*)

---

(*) References marked with an asterisk are available in English translation at the author's website: www.neo-classical-physics.info .